\documentstyle[12pt]{article}
\begin{document}

\noindent Stockholm\\
USITP 05-3\\
September 2005\\

\vspace{1cm}

\begin{center}

{\Large GEOMETRICAL STATISTICS ---

\

CLASSICAL AND QUANTUM}\footnote{Talk at 
the third V\"axj\"o conference on Quantum Theory: Reconsideration of 
foundations, June 2005.} 

\vspace{1cm}

{\large Ingemar Bengtsson}\footnote{Email address: ingemar@physto.se. 
Supported by VR.}

\

{\sl Stockholm University, Alba Nova\\
Fysikum\\
S-106 91 Stockholm, Sweden}

\vspace{8mm}

{\bf Abstract}

\end{center}

\vspace{5mm}

\noindent This is a review of the ideas behind the Fisher--Rao metric on 
classical probability distributions, and how they generalize to metrics 
on density matrices. As is well known, the unique Fisher--Rao metric then 
becomes a large family of monotone metrics. Finally I focus on the 
Bures--Uhlmann metric, and discuss a recent result that connects the 
geometric operator mean to a geodesic billiard on the set of density 
matrices.

\newpage

{\bf 1. Classical preliminaries}

\vspace{5mm}

\noindent How far apart are two probability distributions? An answer to 
this question should encapsulate the ease with which the two distributions 
can be distinguished from each other by means of some kind of sampling. The 
answer will depend very much on what the precise rules of this game are. 

What kind of Riemannian metric is appropriate on a set of classical 
probability distributions over $N$ events, that is on a probability 
simplex? One way of answering the second question proceeds by taking 
the first question seriously, and formulating it for a large number of 
samplings, so that only infinitesimal distances need to be considered. 
In effect, with some minor changes, we will use the idea that long ago led 
investigators in colour theory to introduce a metric on the convex set 
of colours. This is, in a way, very close to the original ideas of Riemann 
himself \cite{Riemann}. But we can also approach the question in the spirit 
of Felix Klein, and first ask what transformations of the 
simplex that we are going to consider. It is encouraging that both 
avenues lead to the same answer. In quantum mechanics---or more appropriately 
quantum probability theory---the situation is more complex, as we will see.     

Let us first deal with the question as Riemann might have dealt with it: let 
us assume that we perform 
${\cal N}$ samplings from a probability distribution $P$ over $N$ 
mutually exclusive events. We were taught by a member of the Bernoulli 
family that if the number of samplings 
is large, then the probability that we obtain the frequency distribution 
$F$ will be   
 
\begin{equation} {\cal P}(F) \sim \exp{\left( -\frac{\cal N}{2}\sum_{i=1}^N 
\frac{(f^i - p^i)^2}{p^i}\right)} \ . \label{1} \end{equation}

\noindent (I have placed the index on the probability vector upstairs, 
because I will be doing differential geometry very soon.) For the moment, 
the point about this famous result is that the width of the distribution 
${\cal P}$ depends on where we are in the probability 
simplex. Let us fix $N = 3$ for definiteness, in which case the set of 
probability distributions is a two dimensional simplex. Bernoulli's results 
provides us with an error ellipse at each point in the simplex. In the 
spirit of Riemannian geometry we now imagine that the simplex is made of 
rubber, and we try to deform it so that all error ellipses become circles 
of a certain standard size. If we succeed we will be looking at a curved 
surface with a definite geometry determined by Bernoulli's theorem. 

It is even easier to do this with equations. The error ellipses are such 
that we are looking at the quadratic form

\begin{equation} ds^2 = \frac{1}{4}\sum_{i,j}g_{ij}dp^idp^j = \frac{1}{4}
\sum_i\frac{dp^idp^i}{p^i} \ . \end{equation}

\noindent This equation defines a Riemannian metric $g_{ij}$, known in this 
context as the Fisher--Rao metric. The factor $1/4$ in front is for later 
agreement with the conventions of quantum mechanics, and is not important 
at this stage. We refer to the tensor

\begin{equation} g_{ij} = \frac{1}{4}\frac{\delta_{ij}}{p^i} \ , \hspace{6mm} 
\sum_i p^i = 1 \ , \hspace{5mm} p^i \geq 0 \ , \end{equation}

\noindent as the Fisher--Rao metric. Up to normalization, it is the Fisher 
information matrix \cite{Fisher}; it was first treated as a Riemannian 
metric by Rao \cite{Rao}. Because of its origins it has a definite operational 
significance, in particular its inverse $g^{ij}$ limits the variance of an 
unbiased estimator according to the Cram\'er--Rao inquality \cite{Cramer}. 
The Fisher--Rao metric can also be used to define a natural 
prior, known as Jeffrey's prior, on the space of probability distributions. 
The point is that any Riemannian metric determines a natural measure which 
is proportional to the square root of the determinant of the metric tensor. 

But what is this metric? We can use the rules for how Riemannian metrics transform 
under coordinate changes to answer this. Define

\begin{equation} x^i = \sqrt{p^i} \ . \end{equation}

\noindent In these coordinates 

\begin{equation} ds^2 = \sum_idx^idx^i \ , \hspace{6mm} \sum_ix^ix^i = 1 \ , 
\hspace{5mm} x^i \geq 0 \ . \end{equation}

\noindent This is recognizable as the metric on a sphere, or more precisely 
on the positive (hyper-) octant of a sphere. So now we know what we have to do 
to our rubber simplex in order to make the error ellipses look like circles of 
a standard size: we have to wrap it on the octant of a round sphere. 

We can use the Fisher--Rao notion of distance to write down an equation 
for the geodesic (shortest) distance between an arbitrary pair of probability 
distributions, namely  

\begin{equation} \cos{D_{\rm FR}(P,Q)} = \sum_i\sqrt{p^iq^i} \ . \end{equation}

\noindent At this point a cautionary note must be sounded: although a Riemannian 
metric can always be used to define a finite geodesic distance between any 
pair of points in space, there is no obvious operational significance 
attached to this here. It is the infinitesimal distance between nearby 
probability distributions that has a meaning, in the limit of a large 
number of samplings---but even so the finite geodesic distance may be a 
useful crutch to study the infinitesimal distance.  

There is a different train of arguments that leads to the same metric. 
Let us define a stochastic matrix as a matrix with positive elements that 
obey $\sum_iT_{ij} = 1$. This is the most general matrix such that $\vec{q} 
= T\vec{p}$ is a probability distribution whenever $\vec{p}$ is. The set of 
stochastic matrices do not form a group, but they do form a semigroup, and 
it is natural to require that any meaningful notion of distance between 
probability distributions should obey

\begin{equation} D(TP, TQ) \leq D(P, Q) \ . \end{equation}

\noindent The idea is that stochastic transformations can only decrease 
the distinguishability of two given probability distributions. Hence their 
mutual distance should decrease, or at best stay constant. A Riemannian 
metric whose geodesic distance has this property is said to be a metric 
monotone under stochastic maps, or a monotone metric for short. 
Then we can rely on a very clear cut theorem, called Chentsov's theorem: 
there exists one and only one monotone Riemannian metric, and this is the 
Fisher-Rao metric. 

Although the statement of Chentsov's theorem could not be simpler, his proof 
(of uniqueness) is quite difficult \cite{Chentsov}; for an accessible version 
see Campbell \cite{Campbell}. Proving that the Fisher--Rao metric has the 
stated property is fairly easy though, and can be done using nothing more 
involved than the Cauchy--Schwarz inequality. Even better, it is fairly easy to 
see what goes on using pictures. To find a stochastic map that stretches 
the flat simplex (i.e., to prove that the flat metric is not monotone under 
stochastic maps), consider the stochastic matrix

\begin{equation} T = \left[ \begin{array}{ccc} 1 & 0 & 0 \\ 0 & 1 & 1 \end{array}
\right] \ . \end{equation}

\noindent This represents a coarse graining of the outcomes of the samplings. 
Such a matrix maps the entire probability simplex onto one of its edges. Now 
it is easy to see what the flow of this transformation looks like. On the flat 
simplex, all points along any line parallel to a particular edge will be taken 
to the same point on another edge, and as a result distances between different 
points may be stretched. The flat metric is not monotone. On the round octant, 
the parallel lines turn into latitude circles. All points on a given latitude 
circle will be taken to a point on a given longitude quarter circle (representing 
an edge of the simplex). Distances between different points 
will decrease if they have different longitudes to start with, otherwise they 
will stay constant. Therefore the round metric is monotone. 

\vspace{1cm}

{\bf 2. Monotone metrics in the quantum case}

\vspace{5mm}

\noindent In the quantum case we do not have probability distributions any 
more, we have density matrices instead. A density matrix is an object that 
stands ready to produce a probability distribution once a POVM is given, but 
it is not in itself a probability distribution. Therefore the statistical 
geometry of quantum states is much subtler than it was in the classical case. 

The easy way to generalize the classical treatment is to focus on stochastic 
maps. A stochastic map of the set of density matrices must take density matrices 
to density matrices, but---as is well known---a somewhat stronger requirement 
is natural. It must do so also when applied to only one of the factors 
of a tensor product. This may sound like a triviality---classically it is 
trivial---but it is not a trivial requirement in quantum theory, and in fact 
it makes the set of maps to be considered much more manageable than it would 
otherwise have been. Thus a stochastic quantum map is defined to be a 
completely positive trace preserving map. While this is a manageable set 
of maps, it turns out that it does not single out a unique monotone 
Riemannian metric. The situation is fully described by a theorem that was 
proved (in stages) by Morozova and Chentsov \cite{MP}, and Petz \cite{Petz}. It 
states that a Riemannian metric on the space of density matrices is monotone 
under stochastic maps if and only if, at a point where $\rho = 
\mbox{diag}(\lambda_1, \dots, \lambda_N)$, it is of the form

\begin{equation} ds^2 = \frac{1}{4}\left[ \sum_{i=1}^N\frac{d\rho_{ii}^2}
{\lambda_i} + 2\sum_{i<j}\frac{|d\rho_{ij}|^2}{\lambda_jf(\lambda_i/\lambda_j)} 
\right] \ . \label{9} \end{equation}

\noindent It is assumed that the function $f$ obeys three conditions, namely

\

i) It is operator monotone.

ii) $f(1/t) = f(t)/t$.

iii) $f(1) = 1$. 

\

\noindent The third condition ensures that the metric is regular at the 
origin. The second condition ensures that the metric is symmetric in the 
eigenvalues. The meaning of the first condition is that 

\begin{equation} A > B \hspace{5mm} \Rightarrow \hspace{5mm} 
f(A) > f(B) \ , \end{equation}

\noindent where $A$ and $B$ are positive matrices of an arbitrary size; since 
positive matrices can be diagonalized it is easy to define $f(A)$. The ordering 
$A > B$ means that $A - B$ is positive definite. An operator monotone function 
is always monotone (in the ordinary sense, that is for $1 \times 1$ matrices), 
but the converse is not true---seemingly innocent functions like $t^2$ 
and $e^t$ are not operator monotone. Nevertheless there exists infinitely 
many functions that obey the three conditions stated, and hence infinitely 
many monotone metrics on the space of density matrices. But all of them share 
the same diagonal part, so we recover the uniqueness of the Fisher--Rao 
metric on the set of classical probability distributions over $N$ events. 

The conditions i)--iii) do not come out of thin air. They appear also in the 
definition of something known as the mean $A\# B$ of two positive operators 
$A$ and $B$ \cite{Ando}. This can be characterized axiomatically, as follows: 

\

a) $A\# A = A$.

b) $(\alpha A)\# (\alpha B) = \alpha (A\# B)$, $\alpha \in {\bf R}$. 

c) $A \geq C$ and $B \geq D$ imply $A\# B \geq C\# D$. 

d) $(UAU^{\dagger})\# (UBU^{\dagger}) \geq U(A\# B)U^{\dagger}$. 

\

\noindent Here $U$ is a unitary matrix. Clearly anything that deserves 
the name ``mean'' must obey this. When a modest continuity requirement is 
added these conditions determine any operator mean to be 

\begin{equation} A\# B = \sqrt{A}f\left(\frac{1}{\sqrt{A}}B\frac{1}{\sqrt{A}}
\right) \sqrt{A} \ , \label{11} \end{equation}

\noindent where the function $f$ obeys conditions i) and iii). Moreover 
the mean will be symmetric, $A\# B = B\# A$, if and only if $f$ 
obeys condition ii).  (You may now begin to suspect that nobody knows how 
to define the mean of three operators. This is correct, as far as I know.) 

Let us give three examples of operator means, and hence of functions obeying 
conditions i)--iii). We choose the arithmetic mean, given by $f = (1+t)/2$, 
the geometric mean given by $f = \sqrt{t}$, and the harmonic mean given by      
$f = 2t/(1+t)$. It is known that the arithmetic mean is the maximal of all 
possible means, while the harmonic mean is the minimal mean. So we get 
two choices of $f$, and hence of monotone quantum metrics, that are 
clearly distinguished in some sense. The geometric mean will also 
be of interest to us, so let us write down the geometric mean of 
two operators explicitly:

\begin{equation} A\# B_{\rm geom} = \sqrt{A}\sqrt{\frac{1}{\sqrt{A}}B\frac{1}
{\sqrt{A}}}\sqrt{A} \ . \end{equation}

\noindent It is clear that if the two matrices commute this reduces to an 
ordinary geometric mean. It is also clear that if they do not commute, 
this is a computational nightmare.  

It is worth noting that, of the three examples that we quoted, two have 
the property that $f(0) = 0$. Inspection of expression (\ref{9}) for the metric 
shows that this means that the metric will diverge at the boundary of the 
set of density matrices, in particular for all pure states. From the point 
of view of statistical distinguishability this is in fact not as absurd as 
it seems at first sight, but I will not go into this here.

\vspace{1cm}

{\bf 3. The Bures--Uhlmann metric}

\vspace{5mm}

\noindent Let us now think of metrics on the space of density matrices 
from a completely different angle. Monotonicity apart, are there any 
other geometrically natural ways to introduce such metrics? On the space 
of pure states, the answer is yes: some very different considerations lead 
to the Fubini--Study metric there \cite{Fubini, Study}. Then the geodesic 
distance between state vectors is given by 

\begin{equation} \cos^2{D_{\rm FS}} = \frac{|\langle \psi|\phi \rangle |^2}
{\langle \psi |\psi \rangle \langle \phi |\phi \rangle } \ . \end{equation} 

\noindent This is a distance defined on the space of physically inequivalent 
state vectors, that is on complex projective space. 
When the dimension $N$ of 
Hilbert space equals 2, it is a sphere of radius $1/2$, but in 
general it is a more complicated space. The point, for the moment, is that 
the Fubini--Study metric arises naturally in a setting that has nothing to 
do with statistical distance. It is a fibre bundle construction. To see this 
we first normalize all vectors in Hilbert space, that is we start with an odd 
dimensional sphere ${\bf S}^{2N-1}$ and equip it with its natural round 
metric. Then we observe that such spheres contain a space filling 
congruence of circles (linked circles in the especially famous ${\bf S}^3$ 
case). These are the fibres of the bundle, and the 
base manifold is the set of all these circles. If we let the shortest 
distance between a pair of fibres define the distance between a pair of 
points in the base manifold, we arrive precisely at the Fubini--Study metric 
on complex projective space. In brief, ${\bf S}^{2N-1}/{\bf S}^1 = 
{\bf CP}^{N-1}$. 

Can this approach be generalized to mixed states? In fact there is a 
setting in which the set of density matrices 
acting on the Hilbert space ${\cal H}^N$ does appear as the base manifold 
of something that is---almost---a fibre bundle. This is so because a mixed 
state $\rho$ acting on ${\cal H}_N$ can always be purified, that is one can always 
find a pure state acting on ${\cal H}_N\otimes {\cal H}_N^\prime $ such that 
the partial trace over the second subsystem is the 
original state $\rho$. It is convenient to identify this composite Hilbert 
space with the set of matrices  ${\cal B}({\cal H}_N)$ acting on 
${\cal H}_N$. That every state can be purified means that we can always 
find an $N\times N$ complex matrix $A$ such that 

\begin{equation} \rho = AA^{\dagger} \ . \end{equation}

\noindent The set of complex matrices is a Hilbert space, with the natural 
scalar product

\begin{equation} \langle A|B\rangle = \mbox{Tr}BA^{\dagger} \ . \end{equation}

\noindent Now a density matrix is a positive matrix of unit trace. It is seen 
that the unit sphere in ${\cal B}({\cal H}_N)$ projects down to the set 
of density matrices under 

\begin{equation} \Pi \ : \hspace{6mm} A \rightarrow \rho = AA^{\dagger} \ . 
\end{equation}

\noindent Namely 

\begin{equation} ||A||^2 = \mbox{Tr}AA^{\dagger} = 1 \hspace{5mm} 
\Rightarrow \hspace{5mm} \mbox{Tr}\rho = 1 \ . \end{equation}

\noindent We also see that the unitary group acts on the fibres, in the sense 
that $A$ and $AU$ will project onto the same $\rho$ provided that $U$ is a unitary 
matrix. Mathematically, this setting is almost exactly the same as that which 
gives rise to the Fubini--Study metric, the only catch being that---in the 
density matrix case---the fibres are not isomorphic to each other but depend 
on the rank of the density matrix. This is 
not too serious a problem, in fact we may ignore it.

Following Uhlmann \cite{Uhlmann} (and, originally, Bures \cite{Bures}) we can now 
define the distance between two density matrices, that is on the base 
manifold of the purification bundle, through

\begin{equation} \cos{D_{BU}} \equiv \mbox{max}\frac{1}{2}(A_1A_2^{\dagger} + A_2
A_1^{\dagger}) = \mbox{Tr}\sqrt{\sqrt{\rho_2}\rho_1\sqrt{\rho_2}} \equiv 
\sqrt{F(\rho_1, \rho_2)} \ . \end{equation}

\noindent The maximum is to be taken over all purifications of $\rho_1$ and 
$\rho_2$ that lie on the unit sphere in ${\cal B}({\cal H}_N)$, so the 
definition of the Bures--Uhlmann distance simply says that the distance 
between a pair of density matrices is the length of the shortest geodesic, 
in the unit sphere, between their two fibres. (It is not particularly difficult 
to find this maximum, provided that one performs polar decompositions of 
every matrix in sight.) The function $F$ is known as the fidelity---it is a 
symmetric function of its two arguments, although this is not immediately 
apparent from the explicit expression. 

Now it is not too difficult to see that if we set $\rho_2 = \rho_1 + 
d\rho$, and choose a diagonal form for $\rho_1$, then the Bures--Uhlmann 
distance is consistent with one of the monotone metrics given above, 
namely for the distinguished choice $f(t) = (1+t)/2$, that is for the 
arithmetic mean. This metric is known as the Bures metric.  

Although we had some difficulties in defining our fibre bundle over the 
boundary of the set of density matrices, it is nevertheless easy to check 
that for a pair of pure states the Bures--Uhlmann distance coincides with 
the Fubini--Study distance. When $N = 2$, that is for qubits, we can 
parametrize an arbitrary density matrix as 

\begin{equation} \rho = \frac{1}{2} \left[ \begin{array}{cc} 1+z & x-iy \\
x+iy & 1-z \end{array} \right] \ . \end{equation}

\noindent The Bures metric becomes 

\begin{equation} ds^2 = \frac{1}{4} \left( dx^2 + dy^2 + dz^2 + 
\frac{(xdx + ydy + zdz)^2}{1 - x^2 - y^2 - z^2}\right)^2 \ . \end{equation}

\noindent This is recognizable as the metric on the ``upper'' hemisphere of 
a 3-dimensional sphere of radius $1/2$, in orthographic coordinates. The 
maximally mixed state sits at the North Pole and the pure states at the 
equator, and states of equal purity have the same latitude \cite{Hubner}.  

Some features of the qubit example do generalize to 
higher $N$. For all $N$ we do have a reasonable amount of control 
over geodesics with respect to the Bures metric, and the pure states form a 
totally geodesic subset. If one performs a horizontal lift of a geodesic 
between a pair of density matrices one finds that the corresponding 
points in the bundle are connected by 

\begin{equation} A_2 = \frac{1}{\sqrt{\rho_1}}\sqrt{\sqrt{\rho_1}
\rho_2\sqrt{\rho_1}}\frac{1}{\sqrt{\rho_1}}A_1 \ . \end{equation}

\noindent Thus $A_1^{\dagger}A_2$ is a positive matrix, and its trace takes 
the maximal value $\sqrt{F(\rho_1, \rho_2)}$. Meanwhile, note 
that here we encounter the geometric mean---of $\rho_1^{-1}$ and 
$\rho_2$---for the second time. A third time is coming. 

The qubit example is misleadingly simple too. 
The Bures geometry does not have constant curvature for higher $N$. In 
fact there are curvature singularities whenever the rank of the density 
matrices drops with 2 \cite{Dittman}. And it is only for $N = 2$ that a 
manageable expression 
for the metric, using the matrix elements as coordinates, is known.      

\vspace{1cm}

{\bf 4. The best possible measurement}

\vspace{5mm}
 
\noindent Let us now try to tie the two strands of our story together. 
Let us first recall exactly how a density matrix can be made to yield a 
probability distribution. We have to choose a POVM, that is to say a set 
of positive operators $E_i$ that form a resolution of the identity: 

\begin{equation} {\bf 1} = \sum_iE_i \ , \hspace{10mm} E_i \geq 0 
\ . \end{equation}

\noindent Here the range of $i$ can be anything; in a projective 
measurement the POVM consists of $N$ orthogonal operators, but that is 
a quite special case. It is assumed that every measurement corresponds 
to some POVM. 

Given two density matrices a POVM helps us to 
two probability vectors, with components 

\begin{equation} p_i = \mbox{Tr}E_i\rho_1 \hspace{5mm} \mbox{and} 
\hspace{5mm} q_i = \mbox{Tr}E_i\rho_2 \ . \end{equation}

\noindent These probability vectors govern the statistics of the 
particular measurement that is associated to the given POVM. With the 
measurement, and hence the POVM, kept fixed, we can compute the statistical 
distance between $\rho_1$ and $\rho_2$ as the classical statistical distance 
between $\vec{p}$ and $\vec{q}$. If we vary the POVM, we vary that distance. 
This suggests that we should define the quantum statistical distance $D$ as 

\begin{equation} \cos{D} = \min_{\{ E\} }\sum_i\sqrt{p_iq_i} \ , 
\end{equation}

\noindent where the minimum is taken over all possible POVMs (that is, we 
maximize the distance over all POVMs). This procedure was actually carried 
out by Fuchs and Caves \cite{Fuchs}, and the answer is precisely the 
Bures--Uhlmann distance. In this specific sense then, the Bures--Uhlmann 
distance is the quantum statistical distance. (There are other ways of 
defining statistical distance, giving other answers, but we will stick 
to this particular definition here.)

The argument offered by Fuchs and Caves is fully explicit, in the sense 
that the best possible POVM is found explicitly. It is a projective 
measurement, associated to the Hermitian operator 

\begin{equation} M(\rho_1, \rho_2) = \frac{1}{\sqrt{\rho_1}}
\sqrt{\sqrt{\rho_1}\rho_2
\sqrt{\rho_1}}\frac{1}{\sqrt{\rho_1}} \ . \end{equation}

\noindent In the specific sense of the argument, this is the best 
possible measurement that one can perform with the aim of distinguishing 
$\rho_1$ and $\rho_2$. The proof is valid as long as one of the density 
matrices is invertible. For a pair of pure states the best possible 
measurement is not unique, but the statistical distance $D$ remains 
well defined---and equal to the Fubini--Study distance between the 
states.

Note that the operator $M$ is precisely the geometric mean of the operators 
$\rho_1^{-1}$ and $\rho_2$. From the symmetry of the geometric mean it 
follows that $M(\rho_1, \rho_2) = M^{-1}(\rho_2, \rho_1)$. Hence we get 
the same eigenbasis, and the same projective measurement, regardless of 
the order in which we consider the two density matrices. Note also that 
the operator $M$ was used in the description of geodesics with respect to 
the Bures metric, and note the simple relation $\rho_2 = M\rho_1M$.  

\vspace{1cm}

{\bf 5. The Uhlmann billiard}

\vspace{5mm}

\noindent Curiously, the operator $M$ has now appeared at three points in 
our story. This is intriguing. At the same time this is awkward, because it 
is a difficult operator to handle. Is there some easy geometric way 
to determine the best possible measurement, given the two density 
matrices to be distinguished? 

Let us see how everything works out for a qubit, where the state space 
is just a round hemisphere (of the 3-sphere), a geodesic is an arc of 
a great circle on the sphere, and a projective measurement can be 
represented by a pair of antipodal points on the equator (representing 
the eigenstates of $M$, in our case). We would like to develop enough 
intuition so that, if we are told the location of $\rho_1$ and $\rho_2$, 
we can point to the pair of points representing $M$ 
without performing any calculations. But before we do so, let us see 
how badly the uniqueness of the best measurement fails if the two 
density matrices to be distinguished are pure. Let us think of the 
two pure states as two points on the boundary of a circular disk, 
subtending an angle $\theta < \pi$. Let a projective measurement define a 
diameter of this disk, making an angle $\theta_A$ with the closest of the 
pair of pure states. A short calculation verifies that the classical 
statistical distance between the states, given such a measurement, 
is $\theta/2 - \theta_A$ if the diameter lies inside the segment 
formed by the states, and it is $\theta/2 $ otherwise. Hence all 
measurements of the latter kind are optimal from this point of view, 
and the quantum statistical distance (as defined above) equals 
$\theta/2$, which is just the Fubini--Study distance between the 
states. This degree of ambiguity in our results will recur in higher 
dimensional Hilbert spaces too.   

When one of the two density matrices is invertible, $M$ is unique. A 
possible guess for $M$ would be that its eigenstates lie on a straight 
line, parallel to the straight line through $\rho_1$ and $\rho_2$ (with 
``straight line'' defined in the sense of convex mixtures). This is the 
sort of thing I mean, but it is wrong. Fortunately a short calculation is 
enough to verify that the eigenstates of $M$ lie on the endpoints 
of the unique Bures--Uhlmann geodesic through $\rho_1$ and $\rho_2$. 
Since geodesics on a round 3-sphere are easy to construct, our program 
to see what $M$ is without calculating it has succeeded---for the qubit. 

For $N > 2$ this picture generalizes as follows. A Bures--Uhlmann geodesic 
in the set of density matrices is always the image of a great circle on 
the unit sphere in Hilbert--Schmidt space, projected down to the base 
manifold. Such a great circle is the intersection of the sphere with 
some real 2--plane

\begin{equation} \{ A: \ A = A_1 + \lambda A_2 \ , \ \ \lambda \in 
{\bf R} \} \ . \end{equation}

\noindent The determinant of such $N$ by $N$ matrices can be zero 
for at most $N$ real values of $\lambda$. But the density matrix to which 
such a matrix projects lies on the boundary of the set of density 
matrices if and only if the matrix has a zero eigenvalue. This is the first 
observation. The second observation is that only rather special great 
circles project down to geodesics. Uhlmann \cite{Uhlmann} pointed out 
that for those special great circles all the roots of the characteristic 
equation are real, which means that the Bures--Uhlmann geodesic actually 
does hit the boundary exactly $N$ times (and then it repeats---the 
great circle covers the geodesic twice). 

Hence we have a geodesic billiard, in which any geodesic bounces from 
$N$ points on the boundary of the set of density matrices. 
(In exceptional cases there may be multiple roots, and then there are 
less than $N$ distinct bounce points, but for now we ignore this.) 
The question is: do these $N$ points have anything to do with 
the $N$ eigenstates of the geometric mean, or more precisely with 
the Fuchs--Caves best measurement operator for a pair of density 
matrices through which the geodesic is drawn?

In the generic case any bounce will occur in the interior of one of 
the maximal faces of the body of density matrices, or equivalently 
at a density matrix with a unique zero eigenvalue. Such a density 
matrix singles out a unique pure state to which it is orthogonal in 
the sense of the Hilbert--Schmidt scalar product. Hence, any geodesic 
through a pair of density matrices $\rho_1$ and $\rho_2$ singles out, 
through its $N$ bounces, $N$ pure states. It has been shown quite 
recently \cite{Asa}\footnote{Actually, 
it was proved only after my talk that this punchline is true. In 
V\"axj\"o, it was just \AA sa Ericsson's guess.} that these 
$N$ pure states are precisely the eigenstates 
of the operator $M(\rho_1, \rho_2)$! It is not 
clear whether this observation gives any special calculational 
advantages, but it paints a nice picture anyway.  

\vspace{1cm}

{\bf 6. Apologia}

\vspace{5mm}

\noindent But why do I tell this story to an audience that is reconsidering 
the foundations of quantum theory? 

The answer is that I think that the story is a beautiful 
illustration of the pre-established harmony between mathematical statistics 
and quantum mechanics. Several other speakers were inclined to refer 
to the latter as ``quantum probability theory''. This would be my position also. 
In quantum probability theory the pure states are always numerous enough 
to form a symplectic manifold. Therefore the name ``quantum mechanics'' 
is a useful one---but still one that misses the main point.
 
\vspace{1cm}

{\bf Acknowledgements:}

\vspace{5mm}

\noindent I thank \AA sa Ericsson and Karol \.Zyczkowski for many 
discussions of these things---and of course Andrei Khrennikov for inviting me 
to visit Sm\aa land, where I enjoyed discussing them with a larger circle 
of friends (as well as swimming in the lakes). For my understanding of 
Bures geodesics I relied on unpublished notes by Hans--J\"urgen Sommers.


\begin{thebibliography}{99}

\bibitem{Riemann} B. Riemann, \"Uber die Hypothesen, welche der Geometrie 
zu Grunde leigen, reprinted in D. E. Smith (ed): {\it A Source Book in 
Mathematics}, vol. two, Dover 1959.

\bibitem{Fisher} R. A. Fisher, Theory of statistical estimation, Proc. 
Cambr. Phil. Soc. {\bf 22} (1925) 700.

\bibitem{Rao} C. R. Rao, Information and accuracy attainable in the 
estimation of statistical parameters, Bull. Calcutta Math. Soc. {\bf 37} 
(1945) 81.

\bibitem{Cramer} H. Cram\'er: {\it Mathematical Methods of Statistics}, 
Almqvist and Wiksells, Uppsala 1945..

\bibitem{Chentsov} N. N. {\v C}encov: {\it Statistical Decision Rules and 
Optimal Inference}, American Mathematical Society, 1982.

\bibitem{Campbell} L. L. Campbell, An extended Cencov characterization of 
the information metric, Proc. Amer. Math. Soc. {\bf 98} (1986) 135.

\bibitem{MP} E. A. Morozova and N. N. {\v C}encov, Markov invariant geometry 
on state manifolds (in Russian), Itogi Nauki i Tehniki {\bf 36} (1990) 69.

\bibitem{Petz} D. Petz, Monotone metrics on matrix spaces, Lin. Alg. Appl. 
{\bf 244} (1996) 81.

\bibitem{Ando} T. Ando, Majorizations and inequalities in matrix theory, 
Lin. Alg. Appl. {\bf 199} (1994) 17.

\bibitem{Fubini} G. Fubini, Sulle metriche definite da una forma Hermitiana, 
Atti Instituto Veneto {\bf 6} (1903) 501. 

\bibitem{Study} E. Study, K\"urzeste Wege in komplexen Gebiet, Math. Annalen 
{\bf 60} (1905) 321.
 
\bibitem{Uhlmann} A. Uhlmann, Density operators as an arena for differential 
geometry, Rep. Math. Phys. {\bf 33} (1993) 253.

\bibitem{Bures} D. J. C. Bures, An extension of Kakutani's theorem on infinite 
product measures to the tensor product of semiinfinite $W_*$ algebras, Trans. 
Am. Math. Soc. {\bf 135} (1969) 199.

\bibitem{Hubner} M. H\"ubner, Explicit computation of the Bures distance for 
density matrices, Phys. Lett. {\bf A163} (1992) 239.

\bibitem{Dittman} J. Dittman, On the Riemannian metric on the space of 
density matrices, Rep. Math. Phys. {\bf 36} (1995) 309.

\bibitem{Fuchs} C. A. Fuchs and C. M. Caves, Mathematical techniques for 
quantum communication theory, Open Sys. Inf. Dyn. {\bf 3} (1995) 345. 

\bibitem{Asa} \AA . Ericsson, Geodesics and the best measurement for 
distinguishing quantum states, arXiv preprint quant-ph/0508133, 2005.

\end{thebibliography}
\end{document}